# Who is Helping Whom? Student Concerns about AI-Teacher Collaboration in Higher Education Classrooms

Bingyi Han, The University of Melbourne, Australia
Simon Coghlan, The University of Melbourne, Australia
George Buchanan, RMIT University, Australia
Dana McKay, RMIT University, Australia

AI's integration into education promises to equip teachers with data-driven insights and intervene in student learning. Despite the intended advancements, there is a lack of understanding of interactions and emerging dynamics in classrooms where various stakeholders including teachers, students, and AI, collaborate. This paper aims to understand how students perceive the implications of AI in Education (AIEd) in terms of classroom collaborative dynamics, especially AI used to observe students and notify teachers to provide targeted help. Using the story completion method, we analyzed narratives from 65 participants, highlighting three challenges: AI decontextualizing of the educational context; AI-teacher cooperation with bias concerns and power disparities; AI's impact on student behavior that further challenges AI's effectiveness. We argue that for effective and ethical AI-facilitated cooperative education, future AIEd design must factor in the situated nature of implementation. Designers must consider the broader nuances of the education context, impacts on multiple stakeholders, dynamics involving these stakeholders, and the interplay among potential consequences for AI systems and stakeholders. It is crucial to understand the values in the situated context, the capacity and limitations of both AI and human for effective cooperation, and any implications to the relevant ecosystem.

**ACM Reference format:**

## 1 INTRODUCTION

Higher education institutions have been eagerly deploying AI-enabled classroom analysis and intervention systems (e.g., [6, 126]) to work with teachers and students in teaching and learning. By collecting, analyzing, and interpreting student learning and performance data [10], AI can continuously observe and assess students' in-class behavior [17, 73]. This analysis can aid educators or instructors with tailored intervention suggestions (e.g., in engineering classrooms [11] and nursing education [31][92]) and offer students real-time personalized feedback [8, 14, 51, 115], thereby enhancing teaching efficiency and learning outcomes. Despite this promise, there is growing apprehension about AI's impacts, such as its effects on privacy and the magnification of existing biases within education [5], which may alter the dynamics in educational settings. However, AI's influence on the cooperative environment and interactions among various stakeholders, including students, teachers, and the AI system itself are not holistically understood.

The CSCW community, with its expertise in navigating complexity among different user groups, is well positioned to shed light on possible future classroom dynamics influenced by AI, to identify problems, and to inform better AI in Education (AIEd) design and adoption.

Existing studies examining the effects of AIEd primarily focus on ethics. These discussions mainly revolve around theoretical and governance-related aspects [119], addressing traditional computational ethical issues such as data use and privacy [15, 75, 132], algorithm bias [84, 96, 161], and algorithm fairness [159]. This relatively narrow ethical focus is observed in relation to AI adoption in general. In the education context, there is a lack of attention given to the interaction between students and AIEd. Students are vulnerable individuals [150] and can bear the direct impact of AIEd, both positive and negative. Although AIEd is explicitly designed to foster positive changes in students' behavior [57], students' perceptions of these potential changes remain unexplored. The way individuals perceive technology can profoundly shape their responses to and engagement with it: a student who views AI as beneficial may interact more openly with it (e.g., [33, 164]), whereas a student who is skeptical of AI might withdraw or disengage from it. A lack of understanding of students' perceptions may undermine successful system adoption [33].

HCI researchers have long recognized that purely conceptual notions of people's interaction with technology might diverge from actual practices. As Suchman [130] and Preece [109] highlight, circumstances around users are ever-evolving and often unanticipated; however, accommodating those unforeseeable contingent consequences is necessary for our designed digital tools. Suchman further posited that the dynamic interaction between humans and their social context constantly reconstructs human practices [130]. Therefore, to fully comprehend the AIEd's implications, we must consider the broader educational ecosystem, encompassing students, teachers, AI, and the educational context – interwoven with social factors that shape how they interact.

AIEd is becoming increasingly complex and pervasive, aiming to revolutionize education [23] While AIEd's intended outcomes are relatively well-defined, many potential unintended consequences remain uncertain. Unlike conventional educational technologies such as laptops or learning platforms like Zoom, AIEd introduces autonomous decision-making systems that pose distinct concerns. For instance, AI systems often rely on copious sensitive data for training models, making them vulnerable to privacy breaches through data leakage, sharing, and hacking [15, 75, 132]. Another key concern involves opaque biases in deep learning AI that are related to race, ethnicity, or gender, and which can unfairly affect student outcomes [18, 79, 159]. Once these harmful unintended consequences arise, they can be persistent and difficult to remediate. Privacy violations, for example, may cause ongoing anxiety for those affected, and biased AI outputs might lead to affected students falling behind in their studies and feeling stigmatized. Although these concerns are recognized to some extent, the full spectrum of potential consequences of AIEd implementation remains largely unexplored.

Yet identifying and addressing negative issues only post-implementation can be unethical and ineffective, especially in sensitive educational settings. The nuances of the educational context and the possible invasiveness of AIEd require a proactive strategy to pre-emptively identify, confront, and mitigate potential unintended consequences. Thus, rather than asking about students' current experiences or exposing them to experimental treatments that could cause both short and long-term harm the nature of which is unclear, we chose to delve further into identifying and preventing potential concerns before they manifest negatively. To this end, we employed a user-centered speculative HCI method with a future perspective – The Story

Completion Method (SCM) – to provoke participants to respond to AIEd scenarios designed based on current proposals from the AIEd community. In story completion tasks, participants are invited to create narratives, which reflect their socially and culturally constructed knowledge and values [74, 155]. The Story Completion Method (SCM) has been effectively used to probe likely responses and evaluate ethical implications regarding emerging or future technology [42].

Our study aims to uncover students' responses to AIEd adoption in higher education classrooms. We pose the following research question: *How do students perceive and respond to AI-assisted analysis and intervention in classroom learning?* Using SCM, we designed story completion tasks with futuristic classroom scenarios. These story prompts are informed by and reflect the advancements and proposals documented in recent AIEd research, as detailed in Section 2 and Section 3. Participants co-created these stories by imagining themselves as studying in such classrooms. By analyzing 65 stories, our study reveals students' concerns regarding AI in the education context, AI cooperating with teachers, and AI interacting with students. We also identified the interactions among various potential impacts of adopting AIEd. We summarize our contributions as follows:

1. Although our study is situated in the context of higher education, students' concerns and ethical perceptions regarding AIEd adoption remain largely unexplored [57], whether in higher education or K-12 education. This gap underscores the value of our work across different educational levels, as it provokes reflection not only in higher education, but also fosters ethics discussions in educational settings overall. Further, higher education students are typically adults (albeit young adults) with more agency than K-12 students. Ethical issues that arise in higher education, where participation in many elements of pedagogy is optional are likely to be more acute with younger, more vulnerable students whose educational participation is often compulsory—we are just beginning to address the ethical challenges wrought by introducing AI into the classroom.
2. We uncover student responses to the newly formed AI-infused classroom learning landscape, especially their concerns and potential impacts on key stakeholders – AI systems, teachers, and students. Our findings shed light on ethical and sociotechnical challenges introduced by AI, to inform better design and adoption of AIEd.
3. We reveal potential interactions between AI interventions and student behavior that are not reported in previous literature, especially the *Situated Changing Circle.* AI monitoring potentially leads to student behavioral change, which in turn can disrupt the AI system's effectiveness. Our findings demonstrate the dynamic nature of AI in education, underscoring Suchman's insights on the situated nature of technology adoption.
4. Our work underscores the significance of viewing AI, teachers, students and the educational environment as an integrated entity, rather than isolating AI within its educational context. This interconnected perspective, highlighting the mutual influence of each element, encourages the CSCW and HCI community to develop holistic, context-sensitive AIEd approaches.

## 2 RELATED WORK

In this section, we first introduce the current approach of AI-assisted classroom analysis and intervention techniques. We then review existing ethical discussions about adopting these AIEd tools.

### 2.1 Integrating AI Analysis and Intervention in Classroom

The AIEd community has made significant effort to design and develop AI applications for higher education institutions, including in four popular areas: assessment and evaluation, profiling and prediction, adaptive personalization systems, and intelligent tutoring systems (ITS) [160]. Many AIEd studies and technological advancement address more than one function area. For example, AI can be used to detect students in-class behavior, assess and predict students learning, diagnose their strengths and weaknesses, and provide customized learning suggestions and teaching strategies. These insights feed into ITS and allow teachers to offer tailored instructions ([11][31][92]), enriching teaching and learning experiences [58].

Recent HCI studies have proposed involving teachers in designing AIEd tools or intelligent systems (e.g., [44, 63]). For instance, methods like the "co-orchestration" model [61, 91, 158] aim to synergize human capabilities and AI strengths, utilizing classroom sensing technologies to inform teachers with data-driven insights in classroom dynamics [58, 77].

Classroom sensing technologies, such as Lumilo [60], Edusense [3], ClassID [106], employ ambient sensors, such as cameras [10, 107], alongside AI techniques like computer vision and machine learning, to detect student movement and behavior, analyze students' engagement, participation, and interactions in educational environments [146, 149]. Specifically, AI-enabled facial and emotion recognition techniques (e.g., [113, 163]) can capture student behavioral data such as head orientation (nodding, shaking, and head poses) [3, 107], body gestures [87], gaze behavior (e.g., eye direction) [4, 64, 83] during lectures or in-class practices [30], as well as facial emotional expressions (boredom, confusion, or frustration) [17, 43]. These data can be fed into learner's affective and physiological state models with indicators of learning performance, such as body motions as learning evidence [87] and arousal in a collaborative learning environment [157]. Derived from quantified indicators [160], AIEd tools interpret students' in-class behavioral performance [10], cognitive-affective status, and real-time emotional and attentional engagement [17, 64, 113].

Based on these data-driven calculations of statistical parameters, students can be categorized such as 'disengaged' or 'struggling' [17, 39, 102] and 'low-performing' or 'at-risk' [101]. These insights about student conditions, performance or drop-out prediction [40, 49, 156] can be delivered to intelligent tutors [113] and human teachers [11, 17, 68] with technology such as TeachActive [11]. This helps to heighten teachers' awareness of students' emotional and cognitive status [131] and assist them in better adapting to learners' emotions [38] or learning progress [39], prompt their reflections on teaching strategies [11, 12, 43, 50], and provide proactive intervention to support students learning [60, 64, 153].

AIEd research has been diligently working on enhancing system functionality [9] through investigating system data and indicators [2] to improve the interpretation of students' interactions and learning patterns in AI-mediated learning environments. A recent review study on AI applications in higher education found that most empirical studies have a quantitative focus [160], and that the current focus often lacks a "pedagogical grounding" [142]. Socio-technical considerations remain understudied in AIEd development [89, 160].

While qualitative empirical studies exist, they typically explore stakeholders' expectations and functional preferences for adopting AIEd [82, 108, 151], concentrating on technical attributes to achieve intended outcomes such as successful integration [65, 103, 108]. Often lacking is an examination of the potential or actual impacts on students [114], especially the unintended consequences and a broader range of ethical risks associated with AIEd adoption [23]. However, the complexities and nuances of educational context warrant urgent critical consideration [122], particularly given the sensitive student characteristics involved, such as gender, sexual identity, and race. This is especially important as students, being a vulnerable group, may be disproportionately affected by these oversights. In the next section, we will briefly review existing concerns about AIEd adoption.

## 2.2 Concerns about Integrating AI in Education

In AIEd, established concerns that are intrinsic to AI techniques dominate the ethical conversation, especially surrounding data use and its computational approaches related to issues [57] such as privacy [143], algorithmic bias and fairness [67, 123], and transparency [112], as evidenced by recent reviews (e.g., [54, 57, 99, 144]). Empirical studies that explore stakeholder perspectives further narrow this scope, with a primary focus on specific themes such as privacy (e.g., [15, 75, 81, 132]) and bias (e.g., [85, 123]). Little attention has been given to the socio-technical impacts on teachers and students stemming from their direct interaction with AI.

This limited (though important) ethical focus has been extended to AIEd governance frameworks (e.g., [110, 120, 128]), which primarily address data-related concerns from a compliance standpoint, seldom considering broader socio-technical implications. This trend persists even in the SHEILA (Supporting Higher Education to Integrate Learning Analytics) framework [141], arguably the most comprehensive learning analytics framework grounded in empirical research. While SHEILA aims to secure stakeholder buy-in and identifies behavioral changes that learning analytics systems should facilitate, it overlooks potential unintended modifications to the learning environment and their impacts on the student body. This gap is acknowledged by SHEILA's architects, with Tsai et al. [140] critiquing the limited research on the impact of technology adoption on the student experience and learning journey. Neglecting these unintended influences risks undermining student trust and acceptance of AIEd, as legal compliance does not necessarily ensure a positive student experience with AIEd systems in practice.

The ethical considerations in AIEd should extend beyond AI and data-centric issues [57] to the real-world context of students-AIEd interactions. While the pedagogical community has raised theoretical critiques about the potential pedagogical impacts of AI-enabled classroom monitoring and intervention, such as affecting learner independence and learning autonomy [134] and the datafication of student behavior and education [90], these arguments often have limited empirical foundation. In contrast, the HCI and CSCW community, with its history of prioritizing ethics in technology design [47] and its emphasis on user-centered approaches [13, 24], advocates for a holistic and empirical understanding of user experience amongst stakeholders—especially, in this case, the students themselves, to truly grasp the nuances of their learning experiences when interacting with AIEd in the learning environment.

Recent HCI and CSCW research has shed light on the unintended socio-technical consequences of emerging technology in education. For instance, a study highlighted that data-driven classroom surveillance [84] can oversimplify children's complex psycho-social behaviors, reducing them to mere data points. Such methodologies risk reinforcing and amplifying existing biases and inequalities in education, particularly those rooted in race, gender, and culture. That study focussed on repercussions from teachers' perspectives. However, it is crucial to recognize that students and teachers might perceive ethical values and impacts differently [158], especially given the inherent power differential between them (e.g., [105]). To date, AIEd discourse has largely sidestepped students' views — a worrisome oversight given students will directly experience AIEd's impacts. Ignoring these concerns could impede ethical and pedagogically useful AIEd adoption and potentially harm the individuals it seeks to benefit.

We have noted that both HCI and CSCW underscore the complex situated nature of technology adoption (e.g., [45, 66]) and the importance of understanding its impacts not just on functionality but also on the broader social and emotional facets of people's lives (e.g., [70]). Yet it is well known that actual user interactions with technology can diverge from what designers anticipate [109, 130]. Indeed, the value promoted by the system might even contradict and undermine design intentions [46]. This is particularly concerning in the educational context, as any potential adverse effects of unintended impacts are significant and possibly harmful. However, as we have stressed, the intricate dynamics between students and the evolving AIEd environment are underexplored.

Our research seeks to bridge this gap in understanding by focusing on the educational context from students' perspectives. While academia and industry are fervently pursuing advancements in AIEd (as described in Sec 2.1), widespread adoption is still in its infancy, with only a handful of learners experiencing its full effects [97]. But given the complexity and potential invasiveness of AIEd, it is crucial to preemptively identify and address its pitfalls before harm is done. To proactively address these concerns, we employ the user-centered speculative Story Completion Method (SCM), which allows us to delve into "what if" scenarios, offering insights that can guide the adoption of more ethically sound AIEd designs.

## 3 RESEARCH METHOD

The overarching goal of this study is to explore students' concerns about AIEd in higher education classrooms. We explore how students perceive classroom collaborative dynamics involving AI, teachers and themselves. Because there has been little consideration of broader sociotechnical implications in current ethical frameworks, we chose not to adopt any pre-existing ethical frameworks to guide our study. Instead, we suggest that our findings can enrich and supplement existing ethical frameworks. The research was approved by The University of Melbourne Human Research Ethics Committee (Ethics ID Number: 24525).

### 3.1 Why the Story Completion Method?

Speculative methods, such as story completion [42], and pre-mortems/post-mortems [145], are widely used to critically evaluate the potential ethical and moral implications of emerging technologies before their widespread implementation [42]. This approach is particularly relevant and ethical in educational contexts, where real-world experimentation might expose students to unintended negative consequences. By using speculative methods, we can proactively explore potential issues in sensitive settings, ensuring ethical considerations are thoroughly examined.

The Story Completion Method (SCM) incorporates elements of speculative design [42, 154] and research fiction [22, 138], inviting participants to create narratives based on provided Story Completion Tasks (SCT). Those story openings help with introducing and constraining the context of our research interest for participants [35]. Stories created by participants present their socially constructed knowledge and values [74, 155], indicating common assumptions about or visions of AIEd's impacts on students and their learning. Recent HCI work has shown the effectiveness of SCM in exploring other sensitive topics (e.g., VR pornography [155] and sex robots [139]), as the third-person storytelling approach with indirectness [35] encourages a more comfortable disclosure of concerns. These qualities make it an excellent candidate for revealing ethical challenges.

XX:7

We acknowledge the important criticism that SCM cannot inform us about actual psychological states [74]. Hence, we interpret our results as potential outcomes ("might be") rather than definitive predictions ("will be"), aiming to uncover students' socially constructed visions of AI intervention in education.

### 3.2 Data Collection

We designed three Story Completion Tasks (SCT), as seen in the Appendix, presenting possible future scenarios of AI-enabled intervention in university classrooms. These scenarios were based on proposed or existing AIEd technologies (Table 1) and were also informed by recent literature-based design fiction studies exploring the impacts of AI adoption in higher education (e.g., [37]). Unlike previous HCI studies exploring future technology design (e.g. [139, 155]) that used brief story openings consisting of only one or two sentences, our SCTs offer more immersive scenarios. Given our study's focus on the situated aspects of AIEd adoption, we provided more descriptions to help participants understand the context. This approach aligns with a recent HCI study [137] that extended SCM with interactive storytelling gamification, further enhancing immersion through SCT.

Scenarios 1 and 2 (Appendix) depict the observation on students' in-class behaviors using AI tools such as Facial Recognition and Emotion Recognition, to help teacher understand students' emotional readiness and cognitive needs for learning so that to provide targeted support. Specifically, Scenario 1 focused on individual observation during lectures, while Scenario 2 emphasized observation during group activities to capture individuals' engagement in more extensive student interactions, so that teachers can better understand the learning progress of the class and adjust teaching strategies. In Scenario 3, the data captured by these systems, including students' engagements and affective states, are utilized for AIEd functions. This scenario described the intended benefits to enhance students' learning with customized learning services, encompassing real-time analysis based on individual education profiles, personalized recommendations on learning pathways and resources, and automatic evaluation, prediction, and interventions.

We acknowledge that our use of story prompts could influence participants' responses. Terms such as "monitoring" might evoke certain narratives. To mitigate this limitation, we clarified in the introductory materials that the purpose of this research was not to cast AIEd technology in a negative light. Rather, we seek to identify perspectives and concerns that can inform the ethical design and improvement of AIEd systems. Our scenarios emphasized the objective of enhancing learning as we tried to maintain a neutral tone to avoid introducing any negative biases.

In addition, we intentionally described the technology artifacts and AIEd functions with a degree of ambiguity, providing only the context, to encourage imagination and elicit a broader range of responses [28]. At the end of each scenario, we guided participants by asking what happens next and how the character would think and behave. Participants were asked to spend 5 to 10 minutes completing each story, to ensure the story quality.

Table 1. Story Prompts Justification – Proposed or Existing AIEd Artifacts

| Scenarios (S) | Functions & Techniques | Code in Story | Examples |
|---|---|---|---|
| S1&2 | Classroom Facial & Emotional Monitoring | [CEM] | [113] |
| In-Class Observation, | Eye Direction / Gaze Tracking during Lecture | [EGT] | [4, 64, 83] |
| Real-Time Evaluation | Awareness & Emotional Engagement Analysis | [AEA] | [43, 59, 60] |

| Scenarios (S) | Functions & Techniques | Code in Story | Examples |
|---|---|---|---|
| | Real-Time Classroom Behaviour Analysis | [RBA] | [3, 17, 83, 107] |
| S3 Intervention, | Automatic Evaluation and Timely Intervention | [ATI] | [9, 17] |
| Prediction, | Prediction and Academic Early Warning | [PAW] | [1, 49, 100, 156] |
| Customization | Adaptive Personalized Recommendation | [APR] | [9, 48] |

[a]Full scenarios are available in the appendix. The functions and techniques in Table 1 are based on proposed or existing AIEd technologies and recent literature. To facilitate referencing, a code has been assigned to each function, marked in the corresponding stories.

We targeted participants recently or currently experiencing higher education (undergraduate, postgraduate, PhDs, and recent graduates). To explore potential cultural influences on the perception of ethical concepts in AIEd, we recruited participants from two different demographics (China and Australia). However, we did not ultimately observe substantive differences between them, and this allows us to conclude that our findings span cultures. Our recruitment did not filter participants based on their majors, level of AIEd experience, or AI literacy, as our focus was on broader concerns encompassing all types of students, considering the diverse backgrounds of students in the actual university contexts.

We collected 71 stories anonymously via the Qualtrics online survey platform. Our survey presented story stems, and collected participant's demographic information including gender, age, and teaching experience, if any. Participants with teaching experience were denoted with a "T" (e.g., P1-T). Six incomplete stories were excluded from our analysis. Of the remaining 65 participants, 54% were male and 46% were female; none identified as non-binary.

### 3.3 Data Analysis

We employed reflexive Thematic Analysis (TA) to analyze the collected stories [26], as it allows us to form an understanding of the overarching shared viewpoints of our participants. The first author led the analysis following Braun and Clarke's [25] guidance. After carefully reviewed all the stories, we coded the stories and organized the codes inductively. These codes were either semantic, which captured the explicit ideas expressed by participants, or latent, which reflected the researcher's interpretations of the underlying meanings present within the data [133]. Then, we sorted the codes into initial themes and sketched their relationships on mind maps [25]. The research team reviewed the mind maps and refined the category structure through multiple rounds. This iterative process adheres to the principles of reflexive TA [27], which advocates for continuous iterative reflexivity and ensures that the analysis and interpretations evolve progressively and recursively [26, 135]. We finally answer our research question, *How do students perceive and respond to AI-assisted intervention in classroom learning,* through three main categories including AI in context, AI and teacher, AI and students themselves. We present those three categories in the following sub-sections, with identified themes highlighted in Bold.

## 4 FINDINGS

This paper is part of an overarching study that aims to understand students' perceptions of AI interventions in higher education classrooms. The overall findings highlight students concerns and expectations across various aspects of AIEd within its adoption context: we have presented the individual dimension in [55], and the pedagogical dimension in [56]. In this paper, we focus on three major concerns regarding the dynamics among AI in the context, teachers, and students: the first is the potential for AIEd's quantification approach to oversimplify intricate educational context and thereby disrupt fairness (Sec 4.1). The second is the collaboration between human teachers and AI, where we observe situated bias concerns and power disparities (Sec 4.2). The third is the fact that students may change their behavior as interactive strategies with AIEd, while these situated responses can intentionally or unintentionally disrupt AIEd's efficiency and effectiveness. (Sec 4.3). Finally, we introduce students' positive perceptions regarding educational processes and approaches (Sec 4.4).

## 4.1 Context: Disruptions to Accuracy and Fairness where AI is Context-Unaware.

Participants' most frequently expressed concern was whether AI can fully represent the actual education context. They felt the **quantification approach** of AI-assisted evaluation **oversimplifies human behavior**. In particular, the nature of human beings and human experience, especially feelings and affective status, are "too delicate to be understood by quantitative indicators" (P36). "The polysemy inherent in human facial expressions adds complexity to their interpretation. Moreover, the connection between these expressions and an individual's inner thoughts remains ambiguous. So, the detection of such nuances by AI systems may be shallow and prone to inaccuracy" (P3). Hence, "the data may not be fully representative of the real-life situation" (P52-T). Here are some examples of **data misrepresentation**:

> "... the system reported that he looked bored, but he was just really tired after working a late shift the night before." (P65-T)
> "The real-time AI analysis detected his lack of confidence for today's lecture. Actually, it was caused by his frustration with his crush who sits nearby. Each glance in her direction intensified his nervousness..." (P5)

The story from P15 above also questions the correlation between the collected attributes of facial expressions and in-class behavior information with actual learning progress. Similar concerns frequently pointed to the quantitative representation's **decontextualizing the learning process and over-digitizing education**. Participants argue that learning outcomes are influenced by a multitude of factors, including individuals' learning motivations and various elements within the lecture environment such as "...the specific content of the lectures, the teaching method of the lecturers and the interactions between students" (P64). The system evaluation, in their view, oversimplifies the complexity of learning situations by solely focusing on surface-level outcomes while neglecting other essential pedagogical factors that shape students' experiences and achievements. An example below emphasizes the limitation of relying solely on quantitative data to comprehend the dynamics and intricacies of individual learning trajectories:

> "The learning curves of most students follow a non-linear pattern, but more like an explosive growth during finals… AI analysis won't know that." (P33)

Another commonly raised concern pertains to the tendency of rigid **one-standard-based** AI systems to **over-generalize educational situations**. These concerns are closely linked to the presence of numerous complex **exceptions** within real-world educational settings, which the mechanistic fact-based nature of AI may struggle to handle in a fair manner. Handling student-related matters in educational settings requires flexible responses. AI systems that rely solely on standardized metrics might neglect the inherent intricacies and diversity within education, and **unfair decision-making** may ensue:

> Exceptional Case: "The student was sick, but he still made an effort to attend class every day and aced his grades. He didn't participate in activities that much and seemed uncomfortable in class, which should've been understandable given the circumstances. But guess what? The system evaluation assigned him a lowest on-class performance score. Talk about unfair!" (P21-T)
>
> Educational Diversity: "…bright students can appear disengaged because they are not challenged enough - which is also a reason for poor attendance… students may have covered some of his subject materials in previous study or may have transferred from another university where this material was covered earlier in the course. These factors all cause variations in student performance." (P59-T)

The diversity within the student population is one obvious example that shows the complexity of education. The stories below described a particular advantaged personality of students when evaluating in-class performance by quantifying interactions and engagements. This mechanism potentially favor extraverts or "expressionists" against introverts, thereby disrupting **fairness**:

> "This student is knowledgeable… He is one of the quieter ones in the class, which is likely why the evaluation system has given him a lower score… this assessment is unfair." (P56-T)
>
> "The system does not consider students' personality and traits that can affect students' patterns of participation in group discussion. While outgoing students who enjoy engaging in discussions may not be affected, those who feel uncomfortable expressing themselves within a group may face difficulties. The system becomes a burden for them, hindering their preferred mode of participation." (P64)

In traditional classrooms, group work may already pose challenges for some students. The examples above demonstrate that AI-based performance evaluation, with its rigid algorithm, tends to be one-size-fits-all that can intensify these challenges for some students in group settings. Thereby, it potentially creates a less inclusive learning environment for diverse personalities.

Achieving fairness, as perceived by participants, necessitates the consideration of all possible situations for students, allowing for compassionate exemption and explanation opportunities. Participants' discontent with the inflexible nature of the AI mechanism and its oversimplification of educational scenarios during assessment often stems from their perception of AI's lack of compassion in decision-making. Particularly, the exceptional cases mentioned earlier were believed to be better handled by the human touch of compassion and subtle empathy. Those attributes are seen as essential in the educational context where AI can fall short, due to its unawareness of the intricacies of education. Interestingly, when participants perceive the inflexible AI evaluation as unfair, they tend to more frequently compare AI and human teachers. Their point revolved around the notion that learning evaluation should regard human factors where teachers' human experience and deliberation are vital to ensure fairness. This point features in the next category.

## 4.2 Teachers: AI Cooperation, Agency, and Power

In this section, we initially demonstrate how participants envisage teacher-AI cooperation and the respective roles of student and teacher in the classroom. We then introduce participants' concerns about, first, bias concerns and, second, power disparities.

*4.2.1 Human-AI Cooperation and Roles.* Participants advocated for integrating human judgment to counterbalance AI's inherent limitation: "AI analysis cannot be seen as the sole evidence or method for evaluating students, as it is unable to understand students' varying situations and circumstances" (P27-T). They see the role of human intervention in AI-mediated learning evaluation, as a critical way to add contextual insight to extend AI's analytical capability. Teachers, adept at discerning intricate classroom dynamics that impact academic outcomes, draw from "their observation of subtle student behaviors and affective states" (P42-T), as well as their understanding of cultural nuances and individual student backgrounds. This depth of teachers' situational awareness empowers them to formulate flexible strategies. The role of AI, therefore, should be positioned as a *supportive complement* to human teachers, furnishing valuable analytical insights without encroaching upon the authority inherently tied to the human facet of education.

Participants often describe a preferred partnership between humans and AI where **human teachers hold agency** while **AI serves as a supportive auxiliary tool**. Stories provide various examples of how human teachers exercise their agency in conjunction with AIEd such as: Customizing the taught curriculum according to AI's analysis, such as adding scientific references based on class progress and student needs; adapting the instructional approach, by drawing on real-time analysis, automated feedback, and suggestions provided by AI; and providing targeted learning support by leveraging AI-generated cues to identify and address students' struggles.

Narratives commonly emphasized the need for **teachers' decision-making authority.** Teachers are expected to actively control the decisions made by AI systems as the analytical insights provided by AI always require **human oversight and interpretation**. One typical comment was:

> "I'm worried if the AI system is given any more weight than what the teacher says. The tool should just only be treated as a reference to assist teaching but should never be given the ultimate say in anything or be the deciding factor in any major decisions (kicking someone out of a course, giving the final grade, marking written pieces of work)." (P60-T)

Especially when conflicting AI analyses occur, teachers are expected to overrule AI. Here are two examples where teachers would aim to ensure evaluation fairness:

> "The professor becomes intrigued by the puzzling difference between the student's mediocre in-class performance, as rated by the AI system, and their impressive exam score. Driven by a strong sense of fairness, the professor carefully investigates how the system evaluates student work. Seeking to understand the truth, the professor kindly invites the student to share their perspective, allowing them to explain their performance. Recognizing their position of authority, the professor considers the option of reevaluating the student's work if it seems necessary." (P45-T)

Stories also mention **teachers' role changing** after introducing AI in education: transforming from observers and evaluators into educational strategists and designers, which requires more enhanced professional pedagogical skills (P13-T). Meanwhile, concerns were raised regarding the potential impact of role transformation on the personalization and individualization of the student-teacher relationship:

> "The experience becomes less personal…Will teachers even bother getting to know him if the machine is taking the attendance and marking the class participation?" (P56-T)
> "The professor tries to remember back over the semester about this student. Before the system was in place he used to take note of students who didn't seem to be engaging, and try to get a feel for what they were thinking. But he has become so used to relying on the system to find these students and offer them a guided path that he no longer has that close contact with his students…" (P65-T)

However, humans might not always ideally lead the Human-AI cooperation. Although the system evaluation might be "ruthless for lacking empathy " (P40-T), some respondents perceived AI as objective and impartial, which is "perfect for playing a role as an **unbiased third party**" (P60-T). They view its fact-based approach as a means to mitigate human biases [125], ensuring fairness in the evaluation process, such as preventing teachers from playing favorites and instead promoting an equitable distribution of attention and learning materials (P67-T). Extending beyond evaluating students, another participant highlights the system's potential to provide a more comprehensive evaluation of teaching compared to student satisfaction surveys, which often reflect only extreme opinions (P65-T). Thus, AI evaluation is considered a valuable tool for a well-rounded assessment for both learning and teaching.

*4.2.2 Bias Concerns.* Nonetheless, AI analysis can be inherently biased [18], as our participants were well aware. Three participants (P30, P59-T, P65-T) express concern regarding unintended biases in AI systems. For example:

> "All automatic systems are only as good as the concepts and algorithms programmed into them… the system may be biased by the ideologies and priorities of the designers. This represents a narrowing of the range of strategies and approaches to which students are exposed. Overall, it is likely that more students will not have their needs addressed." (P59-T)

Participants recognize the potential of AI-provided evidence to address human bias (e.g. teachers' favorites) that may exist in conventional education. Yet, they also express concerns about the **unintended bias** that AI-generated evidence may introduce to teachers' perceptions. Participants highlight the potential **psychological impact on teachers** who may form preconceived opinions from AI analysis before themselves fully assessing students' abilities, akin to a "verdict before trial" (P30). This apprehension is centered around the negative consequences of predictive low grades or ranking, as participants fear that teachers may lose confidence in them (P29) and treat them differently (P43). These examples underscore the potential influence of AI predictions on teachers' biases and perceptions, emphasizing the need to carefully consider the unintended consequences of predictive models.

The empowerment of AI and the use of data then become a shared concern, given those potential implications and unintended consequences. Participants emphasize the unfairness of using past data to predict or prejudge students' future achievements, particularly considering the intricate and ever-evolving nature of the educational context. P30 argues that "even if the predictions are 100% accurate, they remain predictions rather than definitive outcomes." P4 echoes that thought: "Data is power". Further, the reliance of AI systems on historical data can perpetuate existing biases and inequalities present in educational systems. Using past performance as a basis for predictions carries the risk of reinforcing stereotypes and limiting opportunities for students based on their previous outcomes.

*4.2.3 Power Disparities.* Stories also reveal a **power dilemma**: The realization that AI is fallible prompted participants to call for caution and deliberation in leveraging data-driven tools and allocating the power of AI and human teachers. They grapple with questions such as "how much authoritative power should human teachers allocate to AI prediction and analysis when evaluating students?" (P15). We observed earlier that participants place greater trust in human teachers for making final decisions, as they perceive fairness is achieved through discerning special circumstances and possessing contextual understanding and discretion. Hence, teachers are expected to use their authority. However, teachers can be influenced by AI insights, creating a complex dynamic where the power appears to rest in human hands, but can be shaped by the implicit yet potent influence of AIEd; for example, when AIEd directs teachers to a student who is identified as needing more support in a class. Complicating matters further, AI insights themselves may be biased by past data and human designers.

Another example vividly portrays this power dynamic, where a teacher employs AI insights to support (and potentially rationalize) an already-made decision:

> "The Prof doesn't feel the need to increase the scores... This belief is reinforced by the "empirical" data that the Evaluation System provided which he can use as evidence for his decision... He then provided the student with a few suggestions according to the AI-generated report... Ultimately Chris feels more justified in his decision to deny a score increase complaint as it is not solely based on his judgment, but an outside, unbiased, 3rd party." (P60-T)

### 4.3 Students: Behavior Changes Challenge AI Effectiveness

All AI systems rely on data sets that are representative of the interactions they wish to model, yet adopting AI in classrooms will directly impact students' behavior due to their interaction with AIEd. We have demonstrated students' potential behavior changes in detail in [55]. This disruption of behavior means the AI may be based on a model that does not accurately reflect changed circumstances in the classroom, thus reducing the efficacy of the model. This section is not going to demonstrate all possible impacts on individual students, but to reveal how students' potential reactions might further challenge AIEd effectiveness.

The cooperation of AI and teachers in the classroom creates a learning environment perceived by our participants as having ubiquitous monitoring. This perception instills fear in students that their poor performance will be recorded and reported, leading them to believe more unnecessary effort is needed for learning. To meet the perceived stringent requirements for learning and their wish to gain good performance scores, students may adopt interactive strategies and behavior changes, both intentional and unintentional. For example: pretending, acting, and pandering to the surveillance, avoiding making mistakes, being extra cautious of their expression, and forcing active engagement and participation. Detailed impacts on individual dimensions have been presented in [55]. We shall now demonstrate its challenges for AIEd effectiveness.

Most directly, commonly expressed **unintentional unauthentic behaviors, intentional performativity,** and **gaming system behaviors** will all influence the quality of data. However, there are issues that contaminated data will be used to train AI; those altered behaviors are result from AI. These **distorted data inputs** of identification and recognition will further challenge the **technical effectiveness** of the following data-driven functions such as evaluation, customization and intervention. Student participants acknowledged the problems:

> "He pretends that he really understands the newly delivered knowledge. He shows a confident face (Pretending and Acting) … He will work harder at home (Privacy-Seeking) to figure out the hard part rather than allowing the system to detects his struggle (Avoid Making Mistakes) …All the data are generated based on the fact that all students hide their true selves in front of the camera. Those data are pointless." (P52) "Students' reactions were real in the conventional class. Now everyone deliberately presents their perfect self rather than being what they really are (Performativity). The analysis won't reflect the true situation." (P21-T)
>
> "I finally managed to achieve high scores by Gaming the System… The system knows exactly where my study weakness lies, I believe it will reduce the workload for teachers and enhance my learning efficiency by targeting my struggles. Oh wait! My data and information profile are no longer "accurate" starting from this semester, what should I do..." (P18)

One participant pointed out that measures can become poor indicators when they are targeted, referencing Goodhart's Law. It questions the situation where students likely game the system to meet AI expectations and the evaluation will become ineffective:

> "However, as Charlie is a clever university student, he quickly figures out that it is not hard to pretend to be engaged without actually engaging. One can pretend to be reading materials without attempting to understand and one can say many things without actually meaning anything. Not only does he do this, but he also finds out that many other people are doing the same in order to trick the system. A classical phenomenon as described by the Goodhart's Law – 'When a measure becomes a target, it ceases to be a good measure.' Sometimes people have to remember that the system of human behaviours is not invariant with outside observers. The very motion of trying to evaluate people change people. It is hard to think and say how exactly can the AI evaluate without interfering." (P39-T)

Even **hostile behaviors** are expected, with intentionally rebellious strategies to subvert the system's surveillance. For instance, quite a few stories described students' intention to disrupt the accuracy of the system analysis, by deliberately presenting flawed data and "*contaminating the data model*" (P30). These behaviors target **attacking the essential value of AIEd systems** in a covert indirect way. There are also overt direct aggressions such as "...he walked to the front of the classroom and unplugged the power source of the monitoring system with anger" (P2), or even retaliating by attacking the database. Further, stories also describe a mutinous or collective resistance response e.g., "rally classmates to make trouble against the system" (P38)

In addition to disrupting the system outcome, students may also seek to **benefit themselves** by "manipulating the system instead of being manipulated" (P38):

> "The system can benefit him if he uses it well: this professor only read slides in his lecture... He should see those recorded bored faces and improve his teaching. He calls his classmates to exaggerate their facial expressions of boredom..." (P37)
> "He had a strong dislike for algebra. With the system capable of predicting a student's failure, he saw an opportunity to use it to his advantage. If teachers and parents lose confidence in him, he could easily get away with dropping algebra. He began to behave poorly and felt satisfied with his plan." (P48)
> "If AI monitoring tracks my eyeballs, I will deliberately prolong my gaze stare on the illustrations. The analysis might suggest improvements in textbooks by incorporating more pictures..." (P29)

The phenomenon of students engaging in strategic manipulation of the analysis also gives rise to a deliberate **restructuring of student-teacher communication dynamics** with a **new medium**. With the awareness of being under constant surveillance, students become more cautious in their behaviors and expressions, leading to a decrease in direct interaction with both their peers and teachers, particularly when they encounter difficulties in their learning process. Numerous narratives depict students' desire for the system to assist students, particularly introverted individuals who perceive themselves as benefiting the most from this approach. Notably, some stories portray students intentionally gaming the system by amplifying their weaknesses, showing or even feigning struggles, anticipating that the system will prompt teachers to intervene, thereby capturing their attention and securing much-needed assistance. Moreover, once students **trust the efficacy** of the system, they may forego seeking help directly from teachers, relying instead on the system to allocate teacher attention to address their academic needs.

> "This is too challenging for me, but it seems to be easier for my classmates. Perhaps the system can observe this and inform the teacher." (P40)
> "He knows that the system might help the teacher assist him so he might play up his weaknesses to make the AI believe that he needs additional help...Because he is aware of the system tracking his progress, he feels less need to seek direct help from the teacher ...the academic profile will inform the teacher either way." (P60-T)

However, participants frequently voice a **diminished trust** in the accuracy and effectiveness of evaluation across the wider education system. They are skeptical due to their awareness that the system's function is based on "...fabricated data resulting from students pretending and acting" (P45-T); "those inconsistent datasets cannot truly represent the actual learning situation, rendering the analysis untrustworthy" (P52). This distrust extends to the educational system itself, which is perceived to rely on these flawed AI evaluations. For instance, "... the education bases student criticism on quantified method with low fault tolerance, particularly when the system itself might be inaccurate... I will quit the school like this." (P3)

### 4.4  Positive Perceptions: Enhancing Pedagogical Processes

In addition to the positive aspects mentioned in previous sections - such as being perceived as introvert-friendly by providing active support and directing teachers to assist, as an unbiased third party to provide objective evaluations, and creating a new communication medium between students and teachers - participants frequently noted the positive reconstruction on educational process and approaches.

For instance, students foresee AI intervention targeting on individual weaknesses, thereby improving learning effectiveness. Students perceive that AIEd system can provide "undivided, equal attention to each student" (P31), complementing teachers who have "limited time and energy per student... This allows teachers to utilize their energy more effectively" (P36). Additionally, students expect AIEd to assist both teachers and students by providing analytical insights to inform education, thereby increasing the effectiveness of educational process and approach.

### 4.5  Summary

Our findings revealed students' concern that AI's quantification and data representation approach may overlook dynamic and evolving educational contexts to the detriment of teaching and learning (as seen in examples presented in Sec 4.1). This simplifying of learning situations potentially caused by AI evaluating students can also lead to unfairness. Students expressed an expectation for human teachers to use their power and authority to overrule AI, to mitigate oversimplification, and provide contextual knowledge for ensuring fairness (Sec 4.2). We also observed the potential underlying issues of bias and power disparity for AI-human cooperation. However, while students expect teachers to collaborate with AI to ensure fairness, this cooperation in the classroom can lead students to perceive something like a surveillance panopticon. Narratives frequently depicted student behavior changes aimed at either appeasing or resisting AI-powered surveillance. This effect may not only further challenge technical functionalities but also diminish students' trust in the AIEd (Sec 4.3). Students' concerns about the three aspects of AI-mediated education – AI in the educational context, AI and teacher, AI and

student – are often related to ethical issues such as fairness, bias, and power. Nevertheless, students also perceive AIEd can positively enhance pedagogical processes (Sec 4.4). In the next section, we will discuss these perceptions, issues and concerns.

## 5 DISCUSSION

A recent empirical HCI study on AI in future cities [98] highlighted four characteristics of urban intelligence that involve both humans and AI: *Contextual*, emphasizing the integration of knowledge rooted in human understanding of local contexts; *Conscious*, advocating for grounding in real-world experiences and interactions over mere quantification and data reliance; *Collaborative*, highlighting the synergy between tools and information, and between humans and AI; and *Controlled*, emphasizing the importance of human agency in governing AI interactions. As we shall further see below, these characteristics align with our findings on students' concerns about AI's cooperative role in education. Beyond that, our study probes specific challenges with detailed examples. For instance, the quantification approach might risk decontextualizing learning situations (Sec 4.1), and there is an anticipated blending of AI capabilities with human-derived insights, with teachers expected to exert control and make overriding decisions when collaborating with AI tools (Sec 4.2). We further revealed how students' potential behavior change might further challenge AIEd efficacy (Sec 4.3). In this section, we further discuss our findings about the possible effects and challenges of such AIEd.

### 5.1 Understanding the Context of AI Adoption

Our findings reveal student concerns about AI decontextualizing educational situations. To discuss these concerns, in Section 5.1.1, we examine the importance of understanding human values to help AI grasp the situated context, ensuring that the delivered outcomes align with user expectations. In Section 5.1.2, we discuss the challenges associated with contextualizing AI in education.

*5.1.1 Understanding Situated Context and Human Values.* HCI and CSCW research has long emphasized the importance of understanding the situated context of technology use [45] to ensure that its adoption aligns with the social and cultural milieu (e.g., [65, 117, 118]), particularly in high-stakes and sensitive areas [21]. Kaye's [70] exploration of low bandwidth devices for couples in long-distance relationships to communicate intimacy, for example, illustrates how simple devices can foster users' rich interpretations of simple interaction behavior on a single-bit medium of communication, since it is situated in emotionally and socially rich pre-existing relationships.

Such studies remind us to delve deeper into the human values we aim to support, extending far beyond the mere productivity and efficiency that technology could bring. The HCI community has also advocated for solutions that transcend narrow, data-driven approaches (e.g., reflexive AI systems [98]). For instance, the "Whereabouts Clock" [29], which displays family member locations, not only supports family coordination but also contributes to the emotional aspects of family life (though it might also create other problems of its own (e.g., [94])). It provides a compelling example of how technology can enrich lives beyond functional utility. Thus, technology design and adoption should be aware of and responsive to the social and cultural contexts.

Extending these arguments to education, teaching and learning activities are recognized as processes steeped in social, cultural, and emotional capital [152]. Introducing AI into this nuanced educational context goes beyond helping with mere task completion, especially integrating AIEd which relies on quantitative indicators to provide insights on evaluation, teaching strategies, and learning recommendations. Our data reveal the nuanced ways educational experiences might be influenced (e.g., students' performativity) or enriched (e.g., the newly created communication channel between students and teachers). However, the challenge remains in measuring these changes and determining their value within the educational context.

Our findings highlight student concerns that AI itself may oversimplify education and sometimes represent over-digitalization. These arguments from the student perspective in our study align with teacher concerns in a recent HCI study [85] that data-driven classroom surveillance measures codify and simplify the nuanced psycho-social factors that drive students' behavior and performance. Similarly, the education community provides theoretical criticism that behavioral evaluation systems can transform students' learning behavior and its complex social environment into statistical information - a form of pedagogical reductionism [90]. This approach can result in the underrepresentation or inaccurate portrayal of the social context, masking the ambiguity of actual social situations and limiting the interpretation of complexities [121]. Consequently, it decontextualizes [120] and oversimplifies [7] the actual learning process, governing students through classifications and comparisons based on numerical data. However, the potential behavior change observed in our stories may sometimes derail AIEd from its intended objectives.

Such unanticipated consequences (e.g., Suchman [130] and Preece [109]) and value disparities are not new to HCI. An early study on designing education games [46] revealed a discrepancy between the actual and intended values promoted by the designed game, where the design reward structure inadvertently promoted a competitive style of caregiving, rather than the proposed value of cooperation and collaboration. Similarly, smart home technology has been found to inadvertently facilitate control and abuse to intimate partners [94], and electricity information sharing systems can lead to unexpected privacy breaches for household activities [127].

These examples of unintended negative consequences of well-intentioned technologies highlighted the need for extra caution in the educational contexts due to its complexity and the vulnerability of students. Any contradictions between the actual values and intended ones could potentially harm students. Therefore, we must consider the roles and implications of social and emotional aspects of AI intervention for students, teachers, and the educational environment. Understanding potential changes and their impacts can guide us to optimize technology use, align AIEd with students' values and educational practices, and develop more effective and contextually appropriate design solutions.

*5.1.2 Better contextualizing AIEd to Promote Accuracy and Fairness.* A recent HCI study on a student-school assigning algorithm [112] revealed a clash between algorithm modeling assumptions and real-world situations, as the algorithm overlooks the intricacies of existing socioeconomic inequalities. This misalignment can compromise stakeholder values and hinder fair allocation. Similarly, our participants' concerns provide vivid examples of how the educational landscape is littered with complex, multifaceted situations where a simple quantification and data-driven approach could lead to students being unfairly treated. Therefore, a more holistic and context-aware approach to AIEd is essential to ensure the technology does not inadvertently exacerbate existing inequalities or misunderstandings.

Contextualizing AI requires a more comprehensive student data profile [72] including contextual factors [116] such as student background, learning environment, psychological conditions, and individual differences, which can lead to more effective learning experiences.

However, this increased contextualization, achieved through extensive data gathering and analysis, could introduce greater intrusions into the educational process and students' lives, raising issues of surveillance, privacy, and student autonomy [134]. This 'privacy paradox' [143] in education – the inherent tension between full data utilization and privacy – calls for a delicate balance between acceptable data utility and ethical use.

In addition to the need for more nuanced data, the contextual understanding of AIEd should be supported by pedagogical evidence and social theory. AIEd designers should raise awareness of the social dynamics within which the collected data sits, providing social and emotional contexts for data use [85] and making better sense of the digital data [121]. Our findings emphasize student expectations that human involvement will complement AI's limitations in contextual understanding. This resonates with other work on AI-human collaboration and HCI research [84], which stresses the vital roles of teachers in undertaking data work. Such work recontextualizes data to mediate the data-driven educational technology reducing complex students' experiences into simple behavior data. In the next section, we will further discuss approaches that promote ethical and effective human-AI cooperation.

## 5.2 Understanding Human-AI Cooperation

CSCW research has extensively explored human-AI cooperation in high-stakes areas. Similarly to AI in education, collaborative AI in healthcare includes AI assisting radiologists [111] and clinicians [31, 147] with diagnostic insights, enhancing practices for therapists in rehabilitation assessment [80], and collaborative decision-making between humans and AI. These studies highlight the transformative promises of AI, revealing healthcare practitioners' perceptions of AI automating [111] and supporting routine tasks [31], enriching human work by providing richer information artifacts [80, 111], and allowing human professionals to focus on complex intellectual tasks such as "reasoning" and "interpretating" or translating AI results [111] and execution of AI. By synthesizing human-based and algorithmic insights, healthcare professionals perceive AI positively as a collaborative assistant or colleague (e.g., [147] [111]).

In fields beyond healthcare, AI is also seen as a complementary actor in human work, never eliminating the need for human input [148]. Sometimes, AI is perceived not merely as a tool for collaboration but as a subject [162] or an agent with its own agency (e.g., [62, 66]), influencing organizational structures and stakeholders relationships [66]. These insights resonate with our findings where students identified a potential restructuring of student-teacher communication dynamics and teachers' future roles and tasks. Understanding the dynamic and complex roles of humans and AI in its situated context is crucial for configuring effective human-AI partnerships [66].

This effective human-AI collaboration requires strategic insights into task allocation between AI and humans, optimizing interactions, and developing new collaboration paradigms [111]. It has been revealed that human-AI collaborative systems can enhance decision-making accuracy [80] while human and AI can learn from each other's strengths: therapists report greater confidence in their decisions when supported by AI's analysis and feedback, while their comments on AI analysis help improve AI systems. This cooperative dynamic has also been seen in AIEd where AIEd and human can augment each other's interpretations [58, 84]. Despite this, clinicians maintain that AI systems must be verified by human professionals, who remain accountable for the patient treatment outcomes [147]. This resonates with the student preference in our study that teachers hold decision-making authority and overrule AI analysis results to ensure fairness in special cases or pedagogical circumstances. It also resonates with teachers' preferences in another recent study that they should have the final say and take control over AI when AI suggested student pairing plans. These teachers insisted that they know their students' needs better than AI analysis does [77], and that they are responsible for any actions in the classroom when humans and AI share control.

Most importantly, humans must be equipped with an understanding of AI, since the effective cooperation requires a bidirectional exchange of information between humans and AI. It is important to manage expectation in AI and the potential pitfalls of overpromising AI capabilities in human-AI collaboration, especially overhyping the competence of the AI systems [71]. Transparency then becomes critical. For instance, recent CSCW studies ([31] [147]) emphasize transparency and trust in collaborative decision making in healthcare, finding that clinicians desired to understand not only the specific reasoning behind AI decisions (e.g., how the AI processes the inputs and arrives at a decision) but also the basic global properties of the models, including its strengths, theoretical limitations, expertise viewpoints, and design objectives [31]. Enhancing trust involves making AI capability and limitation transparent [147], and providing training for humans for effective use.

*5.2.1 Understanding AI: The Need for Teacher Awareness of AI Bias.* Our participants frequently perceived AI evaluation as an unbiased third party providing objective insight to mitigate teacher preference and ensure fairness, unaware that algorithms, data collection, and interpretation processes can inherently carry and amplify human bias [41, 85]. In turn, this bias can reinforce social power differentials and discrimination [88] and exacerbate inequality in education [85]. This echoes previous studies where people often assume algorithmic decision-making to be fairer and more robust than human judgment [16], perceiving the algorithms as impartial [41]. However, these unintended systematic algorithmic biases are not yet fully understood [18]. While existing studies (e.g., [18]) explore and propose strategies for reducing algorithm bias in education, they often overlook how teachers should manage biased insights when algorithm bias cannot be fully erased, as is often the case. Hence, potential machine biases should be made transparent to teachers. This knowledge can inform collaboration, such as how much to trust AI outputs versus human judgment when they conflict.

*5.2.2 Understanding Humans: Teacher Bias and Attention.* Teacher-AI collaboration needs to be aware not only of the inherent AI bias issues but also of how AI insights can bias human perceptions. Our participants expressed concerns that AIEd analyses might further prejudice teachers by influencing their expectations of students. Unfortunately, these students' concerns are corroborated by teachers' concerns reported in another empirical study [85]. That study revealed that data-driven insights from classroom behavior management systems can reinforce and reproduce teachers' bias, and even institutionalize bias, especially against marginalized groups, perpetuating existing inequality in education. Ironically, we have demonstrated earlier that students often perceive AIEd as an impartial party that can mitigate teacher bias, a perception that could lead to disappointment when confronted with the reality of teachers' perspectives.

There is another striking contrast between students' and teachers' perspectives: Students in our study generally expect that AI can direct teachers' attention to them when they struggle, particularly those who identify as introverts. They view this system as a savior that can help them secure the attention they desire. However, a previous study [85] discovered that teachers often overlook quieter "invisible" students and focus on students with hypervisibility who frequently gain rewards or penalties. Interestingly, some of our participants expressed the view that they would exaggerate their struggles to turn the system into a communication medium, which might actually work to make them noticeable. This, however, raises the question: does this new system encourage students to reshape their behavior and inadvertently penalize quieter students for their natural disposition? The HCI and CSCW communities need to establish measures of potential system bias and raise awareness among teachers about the extent to which they might be influenced by the system if we are to mitigate unintended sociotechnical influences on teachers.

We have discussed how participants acknowledged the potential for AI and human teachers to counteract each other's limitations within a collaboration. HCI and CSCW research must discern how algorithms can be leveraged to mitigate disparities caused by human bias and how human oversight can critically scrutinize and address biases introduced by AI [125], especially biases related to race, gender, sexuality, among others, to prevent injustices. Additionally, it is important to consider other vulnerable groups such as neurodivergent students who may be misread by AI systems. Including their perspectives and needs is crucial to ensure equitable educational technologies.

*5.2.3 Exploring Optimal Frequency of Timely Intervention among Agencies.* While AI intervention, such as shared gaze visualization [131] and "co-orchestration" [61] where AI identifies struggling students in real-time and notifies teachers to direct assistance [69], can be beneficial, it may also create tensions between intervening to help the student and disrupting their learning autonomy [143]. Our previous work [56] reveals that students' perceptions of intervention might vary based on their past achievement: confident students often view intervention negatively as disrupting learning independence, whereas students with less confidence typically see it as beneficial timely support. Given the dynamic and diverse nature of the education system, designers need to consider the possible impacts in each specific situation and the effects of individual student learning styles, preferences and needs [30, 95]. HCI and CSCW researchers need to explore how and when interventions occur among AI, teachers, and students, and determine the frequency that proves most beneficial for learning, while taking care of the bias and inclusivity in the design [96], considering the student population is complex [77], sensitive and vulnerable.

**5.3 Students Behavior Change Challenges AI Effectiveness**

The previous section raised questions about potentially reconstructed students' practices (e.g., to ensure fair attention). As we said earlier when drawing from Suchman's argument [130], the dynamic interaction between humans and technology, situated in their social context, can continually reshape their behavior. In this section, we will discuss how these reshaped human practices can further impact AIEd functions.

*5.3.1 The Potential Impact of Continuous Observation on Students Behavior.* The "Hawthorne Effect" [76], a well-known phenomenon in the field of Human-Computer Interaction (HCI) and psychology, posits that observing a person can result in a change in their behavior. This effect has been considered in the design of many HCI methods (e.g. [20, 78]) - particularly as people tend to behave in a more idealized way under observation, meaning that the observer has to wait for a longer period of time until the monitored people revert to their natural behavior. Our stories show potential behavioral changes (both deliberate and unintended) in student responses to AIEd's observation and evaluation. In the context of AIEd, the observers are obviously not humans but digital agents who will help make decisions. Nonetheless, they may affect behavior. Even if this effect is not guaranteed, its possibility requires careful consideration.

Narratives provided by our participants reveal descriptions of student resistance to the learning environment with AI monitoring. This resistance manifests in students gamifying the evaluation mechanism to achieve higher scores or pass tests, not necessarily to learn but rather to navigate the system and attain their degree. Such gamification and rebellion are seen in other areas where technology is imposed on others by people in authority. For example, children and young adults whose whereabouts are geo-tracked by their parents via smartphone apps, have deployed various clever measures to trick or thwart the digital surveillance, and have poured scorn on both parents and tracking app companies [93].

In our study, projections about students' potential deliberate behavior raises questions about the integrity of the learning process and the potential disconnect between AI-driven evaluations and genuine learning outcomes. AIEd's evaluation approach to quantifying students' performance enables calculations, easier comparisons, classifications, and rankings [36] among students. This facilitates control and governance, where individuals start to reflect on their behaviors and choices [52] and reconstruct their identities by simply accounting for the values [86]. A culture of performativity arises as students adjust their behaviors according to quantifiable measurements of performance success [90]. We saw these potential inauthentic behavior changes of pandering to the evaluation in our participants' stories.

To be sure, the occurrence and permanence of these potential behavioral changes in students as they adapt to AIEd adoption remains unclear. This raises critical questions for the HCI and CSCW communities: Are these changes acceptable? Will they persist in individuals or arise among future student cohorts when and if the technology becomes commonplace? Can AIEd deliver the promised learning outcomes amidst potential disruption of students' learning processes? While adaptive systems have been developed to detect gaming behaviors [19], we must consider the broader implications of these technologies on the educational landscape. This leads us to question the envisioned future of education: Are AIEd systems designed in a way that respects students' autonomy and aligns with their educational values and goals?

Moreover, we lack measures of to what degree students are likely to engage in artificial behaviors to satisfy or undermine the system. Existing AIEd measures primarily focus on system utility and efficiency (e.g., [32]). Future HCI and CSCW work is thus needed to capture the degree of users' involvement in inauthentic behaviors driven by their perception of the system's implicit expectations. For instance, it is essential to understand how much trust students may place in the system's fairness and to what extent they may pretend to exhibit certain behaviors (including dishonestly) to receive positive assessments from the system.

However, here we encounter a dilemma: conducting trials to examine the long-term effects of AIED interventions on student behavior and learning outcomes could be unethical due to potential unintended but harmful impacts. Thus, there should be, we submit, proactive and comprehensive consideration of potential unintended consequences of AIEd before it is designed and implemented.

*5.3.2 Impact of Students Behavior Change on AI Datasets.* What further complicates matters from our stories is this: students' potential changed behaviors, when used as further data inputs to AI systems, could further threaten the accuracy of AIEd analysis and functions. Here, changes in behavior can directly impact data quality since datasets contain noisy, incomplete, misleading, and unrepresentative data. A flawed dataset can undermine the robustness of AI tools [16] and lead to misinformed decisions [144]. Unfortunately, inaccurate analysis based on erroneous data might be treated as valuable evidence for directing students' learning [90] while teachers and institutions accept that as an objective fact [104]. It could thereby create profound negative implications for education and fairness. This issue is akin to challenges in medical AI (e.g., [124]), where flawed data leads to costly misdiagnoses and recommendations. In classroom AIEd, these behavior changes can occur subtly, without students' or teachers' awareness, making it challenging to ensure data authenticity and evaluate dataset quality.

In addition to unintended behavior change, participants' narratives exhibit the possibility of deliberate acts of resistance against AIEd, including sabotaging the datasets. In particular, some stories highlighted students' resistance stemming from their doubts about AIEd's decision-making criteria and perceived as procedural injustice: they view the over-simplified evaluation approach and misrepresentation of data as unfair. This finding resonates with a previous study exploring the relationship between student perceptions of classroom justice and student aggression [34]. The study argues that students' perception of procedural fairness (evaluating the fairness of decision-making processes) positively correlates with motivation and effective learning. To restore equity in the classroom, students may alter or cognitively distort their inputs for being evaluated (e.g. performativity), leave the field (e.g. quitting school), or act against the distributor of evaluation outcomes (e.g. hostile intent towards AIEd system). Our data provides concrete examples of these potential behavior changes, extending the discussion to the classroom learning environment with AI.

Importantly, student resistance, aggression, and hostility toward AIEd, particularly any efforts to disrupt the system by manipulating datasets, may be subtle and difficult to identify. The previous study in the context of traditional classroom [34] suggests that individuals who are currently enrolled in the educational system are more likely to engage in moderate, covert, and indirect forms of aggression towards perceived unfairness, due to the lower risk of negative consequences of such acts. However, in the context of AIEd, students' acts of aggression might be more hidden. In our stories, students' aggression targets the system's core values by passively withholding real information or data during the data collection process. This can manifest in various ways, as shown in our data: faking facial emotions or performing learning behaviors that don't reflect their true engagement of understanding; conducting antisocial strategies such as "contaminating the data model" (P30), or even mutinously rallying classmates to disrupt the system collectively. This covert resistance underscores the impact of student behavior changes on AI dataset, an unanticipated challenge that needs to be considered in the AIEd design and adoption.

When our participant P39-T referred to Goodhart's law by saying, "When a measure becomes a target, it ceases to be a good measure" [129], they highlighted how system can be easily manipulated by students altering their behavior to achieve high scores, thereby misleading evaluation progresses. Similar behavior changes to distort data and meet metrics for accomplishing individual's situated goals have been observed in organizations using analytics for decisions making and assessing worker performance [136]. This organizational study [136] applies signaling theory to describe Goodhart's law: when data produced by workers are used to measure their performance or inform decisions, these data as signals are not neutral. They are often manipulated to convey a specific message or achieve particular objectives. Consequently, measures become ineffective due to the faulty assumption that the signals are honest, reliable, static, or interpreted accurately. The validity of metrics and analytics insights can be thus rendered doubtful.

These signaling acts (either gaming, conforming, or resisting the evaluation mechanism) can potentially empower workers or students, allowing them to assert agency. However, they can also create a feedback loop: technology adoption influences individual behavior, which in turn affects the systems performance—as seen in Fig.1. This interplay raises pressing concerns: an overly invasive system like AIEd might not only bring unintended impacts on students and the pedagogical landscape, but also jeopardize system effectiveness. This concern has been similarly raised in workplace, where, under the influence of monitoring systems, work may become a "strategic, iterative cat-and-mouse game" between organizations attempting to quantify worker performance and workers who continuously adjust their behaviors to manipulate the metrics [136]. Nevertheless, unlike AI used to evaluate work performance, the intended outcome of classroom observation AI is not only to assess students but to help teachers be aware of classroom situations so that to provide targeted assistance to help students to learn.

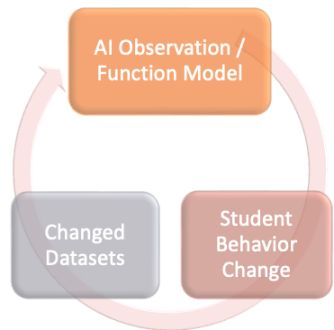

Fig. 1. Recreation of the Repeating Changing Circle.

As HCI and CSCW researchers, we must assess the possible implications on both individuals and ecosystems like education in which AI, students, teachers, and the educational environment are situated, especially when the impacts can be negative, and the outcomes can misalign with the intended objectives.

## 6 LIMITATIONS AND FUTURE WORK

The Story Completion Method (SCM) is a relatively novel method, and the extent to which the framing of the story stems influences participants is not yet fully understood. As noted earlier in the method section, our scenario framing, particularly the use of terms like 'monitoring', may have unintentionally steered responses towards more negative perceptions, eliciting specific emotional and behavioral reactions in narratives. While this approach can help highlight negative feedback and better prepare designers for potential worst-case scenarios, we recognize the importance of careful language use. This insight is valuable for future speculative research, as it emphasizes the need to refine scenario crafting to minimize biases and ensure a diverse array of responses is captured.

As a speculative method, SCM is used to explore commonly held assumptions from a social constructivism perspective. Our study has shown the effectiveness of SCM in detecting a range of possible implications of future technology adoption. The third-person perspective, rather than asking "your perspective", enables participants to project their fears and anticipated behavioral changes when addressing their considerations in sensitive contexts like education. The HCI and CSCW community could leverage this method to understand the potential ethical and educational impacts of other emerging Ed-tech, such as ChatGPT in education, from the viewpoint of students, teachers, and institutions alike. Of course, such speculative methods need to be complemented by studies of actual contexts where that is possible and, furthermore, ethical. Our study focuses on the context of AIEd's adoption in higher education where student groups are university students.

As students in different age groups may interact with and perceive the system differently, future studies could extend the participant age range to include under-18 teenagers or younger children. Additionally, future work should expand to other stakeholder groups such as institutions and educators.

## 7 CONCLUSION

Integrating AI in education transcends mere technological endeavor. It is a complex process deeply rooted in the social, cultural, and emotional landscape of the educational environment, and involving the diverse backgrounds and learning needs of students. Our study explores how students perceive AI in this intricate setting, revealing the dynamic interplay between AI, teachers, students, and the broader educational system. We spotlight critical concerns: AI's potential to decontextualize education settings; AI-teacher collaborations fraught with biases and power disparities; and AI's potential impact on student behavior, which can in turn challenge AI's technical efficacy. Our findings highlight the possibility that changes in one component of this ecosystem can lead to unintended (negative) implications in others. Since education settings are sensitive and students are vulnerable, there is a need for AIEd designs and implementation to consider these contextual nuances. We would therefore re-emphasize to the CSCW and HCI communities the importance of a holistic approach, one that considers AI, teachers, students, and the learning environment as an interconnected whole, rather than isolating AI in a technical vacuum [53]. Such an approach should help ensure that AIEd systems not only enhance learning outcomes but also integrate ethically into the entire educational experience. This balanced consideration of AI's potential alongside a range of concerns paves the way for a more effective, ethical, and contextually sensitive AI-facilitated education.

## APPENDIX
**Story Completion Tasks**
* The codes presented throughout the stories correspond to the functions and techniques outlined in Table 1, which are based on existing research proposals and AI-driven educational tools. Each code, such as [ATI] for Automatic Evaluation and Timely Intervention, is used to reference specific AI functionalities integrated into classroom settings.

**Scenario 1:** During this summer break, Charlie's University has implemented an intelligent AI education system to assist teaching. It is the first lecture of this semester. Charlie walked into the classroom. He heard his classmates whispering: There will be an In-Class Surveillance by using the Classroom Behavior Analysis System:
- When students walk into the classroom, the system records their attendance by Facial Recognition Technology [CEM]. It will then be generated as engagement data stored in the Classroom Attendance System.
- During the class, the system will capture students' affective (emotional) states, such as when a student is either confused or confident, using an Emotion Recognition Technology [CEM]. Then the analysis report will be created for teachers to understand students' emotional readiness and cognitive needs for their learning [AEA][RBA].
- Each student's learning material and feedback will be generated via real-time analysis in class [RBA], to build a profile of the student's progress in each class. Across a term, this data will generate a personal "Education Profile".

Now the class bell rings, Charlie feels …

**Scenario 2:** The lecturer said: "Let's have a group discussion". Charlie then heard another whisper from his classmates:

- During the in-class groupwork practice, the system will also identify what each student is doing and saying, by using Facial Recognition and Voice Recognition. Eye Tracking Technology will be used to capture the amount of time that each student is focused on each learning resource [EGT]. Then the analysis report will be created by Eye Tracking for the teachers to understand the learning progress of the class and the efficacy of the chosen learning recourses.

Now the classroom becomes noisy and full of conversations, Charlie ...

**Scenario 3:** On another day, Charlie learns that there is a digital "Education Profile" that tracks all education-related data for every student (e.g., data of in-class emotion, voice, facial expressions, gesture, academic scores). This profile provides real-time analysis during class and identifies each students' academic weaknesses and strengths [ATI]. Based on this analysis, it suggests personalized learning pathway and generates tailored recommendations for learning resources targeting areas of improvement [APR].

- The Education Profile will work with an Academic Early Warning System [PAW]: It has a high accuracy of predicting when a student is likely to drop-out or fail from a subject. It captures the learner's mindset and how it changes over time, such as a student's attitude towards a lecture changed from positive to negative and unwillingness to engage increased. It has strong indicators to track students' progress and identify changes in each student's learning confidence and motivation toward a course. The system can automatically trigger an alert to ask for timely interventions to support students' learning, such as providing targeted feedback to teachers [ATI].
- The Education Profile will also work with an Intelligent Evaluation System, which automatically assesses homework and provides customized learning assistance when students' struggles are identified [ATI].

Later that day Charlie is in a group exercise on algebra, on a topic he finds quite hard. He knows the Education Profile is going to be used to see how he and his class are doing …. What happens next?